# Novel Tin-based Gel Polymer Electrode and Its Use in Sodium-ion Capacitor

Nitish Yadav, S.A. Hashmi*

*Department of Physics and Astrophysics, University of Delhi, Delhi, 110007, India*

**Abstract**

We present a novel tin based flexible, gel polymer negative electrode for use in sodium-ion devices. The electrodes have been prepared by trapping tin micro-particles in a porous electrolyte membrane formed by immersion precipitation phase-inversion and soaking in a sodium-ion conducting liquid electrolyte. The tin particles are the active electrode materials as well as electron conductors, while the gel polymer enable efficient ion-transfer to the tin particles. The composite gel electrode and its performance in sodium-ion capacitor has been analyzed using scanning electron microscopy. x-ray diffraction, cyclic voltammetry and galvanostatic charge-discharge. The specific capacity of the capacitor shows an initial increase, up to 175 charge-discharge cycles. This increase has been explained by the increase in surface area of tin due to disintegration of the larger particles upon repeated insertion/de-insertion of sodium. A high capacitance of 210 mAh g$^{-1}$ at 1 A g$^{-1}$ has also been measured.

## 1  Introduction

Lithium-ion secondary batteries are the most important energy storage medium with a wide variety of uses from portable electronics to electric vehicles and grid storage. They offer high energy densities coupled with relative safety and durability, but lack the high power and quick charging capabilities required in a number of important applications such as electric vehicles (for quick acceleration or even kinetic energy recovery systems), high power industrial tools, and portable electronics. Additionally, the limited resources of lithium and its erratic geographic distribution pose significant challenges. Supercapacitors, particularly electrical double layer capacitors, have been on the other extreme of the application spectrum, with their high power capabilities non-linear galvanostatic charge-discharge profiles. They also have high cyclic stability (of the order of many hundred thousands of charge-discharge cycles) owing to physical charge-storage mechanism (unlike redox reaction based charge-storage in batteries). Additionally they utilize carbon based electrodes that are easy to manufacture, and pose negligible stability issues. However, they suffer from the problem of lower than needed energy density making them unsuitable for practical use despite the above mentioned advantages. A possible solution to this problem is to prepare hybrid energy storage devices, also known as metal-ion capacitors, that have one battery-like electrode (Faradaic, slow charge-storage), and the opposite electrode EDLC-like (physical, fast charge-storage). Sodium-ion capacitors (NICs) are one such technology, that not only promise to resolve the energy vs. power density imbalance, but are also viable substitutes for comparable lithium-ion based technology. NICs have energy and power densities intermediate to EDLCs and sodium ion batteries [1–4]. The EDLC type electrode in an NIC provides high rate capabilities due to its electrostatic nature of charge-storage and reduces the overall polarization in the cell, while the battery-type electrode offers a high capacity. A number of developments on sodium-ion capacitors have recently been reported. Sodium intercalation materials such as carbon nanotubes [5], expanded graphite [6], hard carbons [7], porous carbons [8], $Na_2Ti_3O_7$ nanorods [1], $Na_3V_2(PO_4)_3$-carbon composites [9], $NiCo_2O_4$ [10], $TiO_2$, Pb, Ge, Phosphorous, Sn, $Fe_2O_3$, CuO, $SnS_2$, $FeS_2$ [11], etc. have been used as negative electrodes, whereas activated carbon [1], two-dimensional vanadium carbide (MXene) [12], $Na_2CoSiO_4$ [13], $\lambda$-$MnO_2$ [14], etc. are used as positive electrodes.

*Corresponding author.
Tel. number: +91-9871088201
Email addresses: sahashmi@physics.du.ac.in, hashmisa2002@yahoo.co.in (S. A. Hashmi).

Tin (Sn) is another important sodium negative electrode with an alloying potential in the range 0 - 0.8V versus Na/$Na^+$. Sn has been reported as negative electrode for NICs and NIBs, in various forms, such as nano-forests [19], alloys [20], and as composite with wood fibers [21]. However, Sn has certain limitations that prevent its utilization in practical devices. Sn undergoes pulverization upon repeated charge-discharge causing structural instability of the electrode and so performance degradation is observed [22,23]. In this work we report a solid state NIC cell prepared using PPE films and a novel tin-gel polymer electrode with an aim to develop a solid-state NIC cell with specific energy and specific power which are intermediate to both batteries and supercapacitors, while having sufficient reversibility of charge-storage mechanism, and cycling stability. The tin (Sn)-gel has been used as the negative electrode and biomass derived activated carbon as the positive electrode. Porous polymer membrane (PPE-Z0) is used as Na-ion conducting electrolyte and separator. The PPE film, has been prepared by activating a porous polymer film, made from a simple phase-inversion method, with liquid electrolyte $NaClO_4$-EC:PC.

## 2 Materials and Methods

*2.1 Materials*

PVDFHPF (average molecular weight ~400,000) was procured from Sigma-Aldrich and used as received. $NaClO_4$, ethylene carbonate (EC) and propylene carbonate (PC) were procured from Sigma-Aldrich. Flexible graphite sheet (~250-mm thick) was purchased from Nickunj Eximp Enterprises., India. Tin powder was procured from Merck.

*2.2 Synthesis of tin-gel polymer and activated carbon electrodes*

Tin (Sn) based gel polymer electrodes for use in sodium-ion capacitors were prepared as follows. First a 1M $NaClO_4$ solution was prepared in EC:PC (1:1 v/v). 0.8 g of the $NaClO_4$-EC:PC solution was taken along with 0.2 g PVDFHFP and dissolved in 10ml of acetone. 2 ml of this new solution was then mixed with 0.071 gm of Sn powder (this amount corresponds to the minimal quantity of Sn in electrode) to make the final electrode-solution. This solution is ultrasonicated for approximately 1 hour to enable proper mixing. This solution is then drop-cast on graphite current collector sheets and allowed to vacuum dry overnight at room temperature to obtain the final electrodes. To prepare the activated carbon electrodes, a slurry of the active material (activated carbon) powder with conducting agent acetylene black (AB) and a binder PVDFHFP in the mass ratio of 70:20:10 was prepared in the solvent NMP. This slurry was then drop-coated on flexible graphite sheets of dimension 1 cm × 1 cm, where they were left at ~70 °C for about 30 minutes to evaporate the NMP. These coated films were dried in vacuum at ~100 vC overnight before use. The mass of the total material on each dried electrode was calculated by measuring the mass difference between the loaded and unloaded electrodes.

*2.3 Preparation of porous electrolyte films*

0.2 g PVDFHFP was dissolved in 1.2 g of NMP and stirred thoroughly on a magnetic stirrer at ~40 °C overnight. ~0.2 ml of this resultant solution was spread on a glass slide of about 3 cm × 2.5 cm dimension and pressed uniformly with another glass slide of similar dimensions. The slides were then separated and quickly and smoothly immersed in a bath of water:NMP (80:20 v/v) mixture. The polymer coated side of each slide was kept facing upwards during this process. Presence of 20 vol.% NMP in the non-solvent solution helped in slowing the non-solvent diffusion into the polymer rich solution leading to better mass transfer along the entire thickness of the film. Phase-inversion happened as the water entered into the solution, and PVDFHFP diffused to polymer rich regions leaving pores with mostly water filled in them. The films were then taken out from the phase-inversion bath and placed in a 100 % water bath for about 2-3 minutes. After this, the films were placed between two pieces of tissue paper to remove excess water from the surface and then transferred to a petridish. The petridish was then kept in vacuum at ~60 °C overnight to remove any residual NMP or water to obtain the dry, phase-inverted membranes. The dried film was then activated by immersing overnight in the liquid electrolyte solution 1M NaClO$_4$ in EC:PC (1:1 v/v).

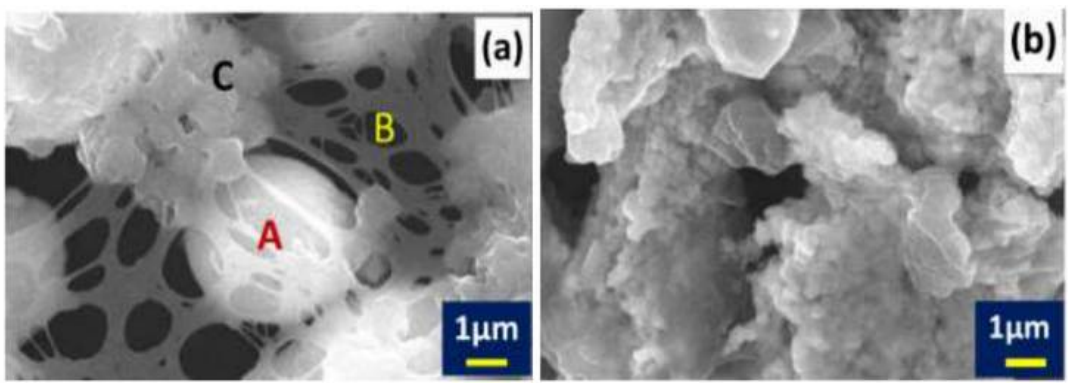

**Figure 1.** Scanning electron micrographs of (a) as prepared Sn-gel polymer electrode (**A**, **B** and **C** denote various regions as explained in text), and (b) same electrode after 30 charge-discharge cycles.

*2.4 Fabrication of NICs*

An activated PPE film was cut slightly larger than the electrodes and sandwiched between a tin-gel electrode and an AC electrode to obtain the solid-state, flexible NIC cell. The masses of the electrodes in the NIC cell were maintained at a fixed ratio to optimize cell performance. The cells were placed in a sealed split test cell (EQ -3ESTC, MTI Corporation, USA) for electrochemical testing. For preparation of 2032 coin cells, an electrode of 1 × 1 cm$^2$ area was used with a larger circular electrode (dia. ∼15 mm) that was coated in a similar fashion.

## 3 Results and discussion

*3.1 SEM studies of Sn-gel electrodes*

Fig. 1(a) shows SEM image of the fresh electrode consisting of Sn-particles dispersed in PVDFHFP/EC:PC/NaClO$_4$ gel polymer. Fig. 1(b) is the image of an electrode from an NIC cell that has undergone 30 charge-discharge cycles. As can be seen in the images, the fresh electrode shows clearly the presence of almost spherical Sn particles (denoted by A) (~5 μm diameter) entangled in a web of the gel-phase (denoted by B). The gel-phase itself gathers/clumps at a few places (denoted by C). Such clumps can act as reservoirs of ions near the Sn particles. Additionally, one can notice that a few Sn particles are also entirely enclosed in an outer covering of the polymer gel-phase, increasing the direct contact with the electrolyte part. Such features are useful for high-rate capabilities, high capacity of the NIC device and for adjusting for the volumetric changes and consequent splitting of the Sn particles upon repeated alloying/de-alloying with sodium. The web like part seen in this figure possibly acts as the mesh holding the Sn particles. Upon undergoing many charge-discharge cycles in an NIC device, the Sn-gel electrode undergoes a substantial structural changes. The polymeric web-like part disappears and is now almost completely replaced by the clump-like polymer phase that almost entirely covers all the Sn particles in the electrode. The Sn particles themselves no longer remain spherical in shape, and appear chipped and split. These images indicate a very useful property of this electrode. Upon repeated cycles, the Sn particles undergo repeated volumetric changes which leads to degradation in the form of splitting and chipping. The gel also undergoes repeated changes which lead to its clumping as seen in the images. As the Sn particles chip and break apart, the flexibility in the structure of the gel allows it to coat these smaller particles, preventing degradation in electrochemical performance of the electrode, as evident in cyclic voltammetry and galvanostatic charge-discharge studies, discussed below.

*3.2 XRD study of fresh and used electrodes*

The X-ray diffraction patterns of the used and fresh electrodes are shown in Fig. 2. The peaks observed here have been compared to those reported recently by Prosini et al. [24]. The peaks visible near 30.5°, 32°, 44°, 45°, 55.5°, 63°, 64°, 64.5°, 72.5°, 73.5° and 79° correspond to that of metallic tin having a tetragonal structure. Non-negligible contamination of $SnO_2$ is evident from the peaks at 26.5° and 54.5°. The peak for tetragonal Sn at 55.5° vanishes altogether after cycling while the other peaks are visible weakened after 30 cycles of charging and discharging. On the other hand, the two peaks corresponding to $SnO_2$ show significant increase in their intensity after 30 cycles indicating that the $SnO_2$ structure not only undergoes mechanical degradation, but also corrodes from irreversible reactions happening during the repeated charge-discharge process. The source of the oxygen could be the degradation of the perchlorate and carbonate moieties from the electrolyte we have chosen and thus the corrosion can be reduced by a suitable choice of electrolyte.

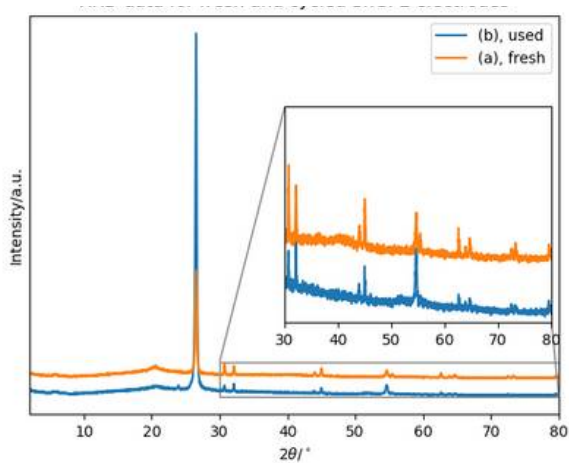

**Figure 2**. XRD patterns of (a) as prepared Sn-gel polymer electrode, and (b) same electrode after 30 charge-discharge cycles.

*3.3 Electrochemical performance of NIC fabricated using Sn-gel polymer electrodes*

Typical cyclic voltammetry curves of the NIC fabricated using the Sn-gel electrode as anode and biomass-derived activated carbon as cathode are shown in Fig. 6.9. The CV studies were carried out between 0 and 2 V where most of the charge-transfer processes happen, at various scan rates. Tin has the ability to alloy with sodium at a low potential (between 0 - 0.8 V vs. Na/Na$^+$) [25] making it a good negative electrode for sodium-ion capacitors and batteries. On the other hand, activated carbon electrode provides double layer formation capability at as high voltages as 3 V vs. Na/Na$^+$ and is only constrained by the stability of the electrolyte used. In accordance with this information, distinct features concerning the alloying of sodium with tin particles can indeed be seen in the low voltage region (0 - 0.5 V) of the CV curves during the cathodic run and corresponding to de-insertion of sodium from tin, providing important contribution to charge-storage in the device. Beyond 0.5 V a more or less rectilinear shape is obtained for the curve, due to the dominance of double layer charge-storage mechanism of the AC electrode. However, a small hump is seen in the anodic region at ~1.4 V indicating alloying of sodium with tin. That the cell can maintain its shape even for high scan rates (500 mV s$^{-1}$), indicates its high rate capability. Overall, CV analysis reveals a reversible movement of sodium-ions into the tin particles, with a large contribution from the Sn gel electrode, providing increase in the charge-storage capacity of the cell, and a large double-layer charge-storage contribution from the AC electrodes is also seen that can maintain efficient charge-storage even at moderately high scan-rates.

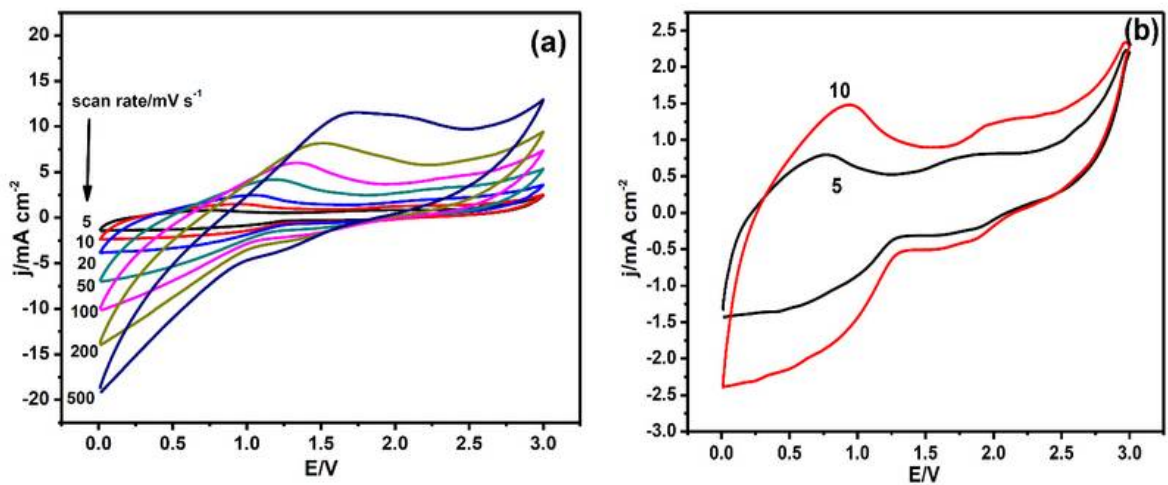

**Figure 3.** (a) Cyclic voltammetry plots at different scan rates (indicated) for sodium-ion capacitors prepared using Sn-gel polymer negative electrodes and biomass-based activated carbon positive electrodes with porous polymer film electrolyte, (b) zoomed CV plots at 5 and 10 mV s$^{-1}$ scan rates.

To further understand the nature of charge-storage and the effectiveness of the Sn-gel electrodes in NIC devices, charge-discharge experiments were carried out at constant applied currents of various magnitudes. For the experiments the cells were cycled between 0 and 3 V. Fig. 4 shows the typical plots of the NIC cells fabricated from Sn-gel and biomass derived AC electrodes. The capacity per unit mass of Sn in the cell is found to be 210 mAh g$^{-1}$ at 1 Ag$^{-1}$ current, while the specific energy and power are found to be 38.6 Wh kg$^{-1}$ and 666.8 W kg$^{-1}$. The charging process is clearly seen to have a complex potential versus current profile, which is not entirely linear and gives the indications of slow Faradaic charge-storage processes with changes in slope observed at about 0.7 V, 1.5 V and 2.5 V at current density of 0.5 mA cm-2 (~1 Ag$^{-1}$). The discharge profiles show two clearly-defined regions, one between 3 V and 1 V, and the other between 1 V and 0 V. The magnitude of the slope for the upper, high-voltage region is comparatively larger than that of the lower voltage region. This higher voltage region is where energy is provided to the external circuit predominantly from the double-layer formed in the AC electrodes and thus has a lower value of capacity as the charge-storage happens on the limited surface of the pores of the carbon. The lower voltage region is where energy is released as sodium-ions are removed from the bulk of the Sn-gel electrode into the electrolyte. This process has a larger capacity due to the large number of sodium-ions involved in alloying/de-alloying with the bulk of the Sn particles. Clearly, the use of Sn-gel electrodes has led to a positive impact on the performance of the NIC device. Also, the small size of the particles, embedded in the polymer matrix, along with the subsequent restructuring of the electrode microstructure, as seen in the FESEM images, is able to maintain the high current operations and high capacity even after a large number of cycles.

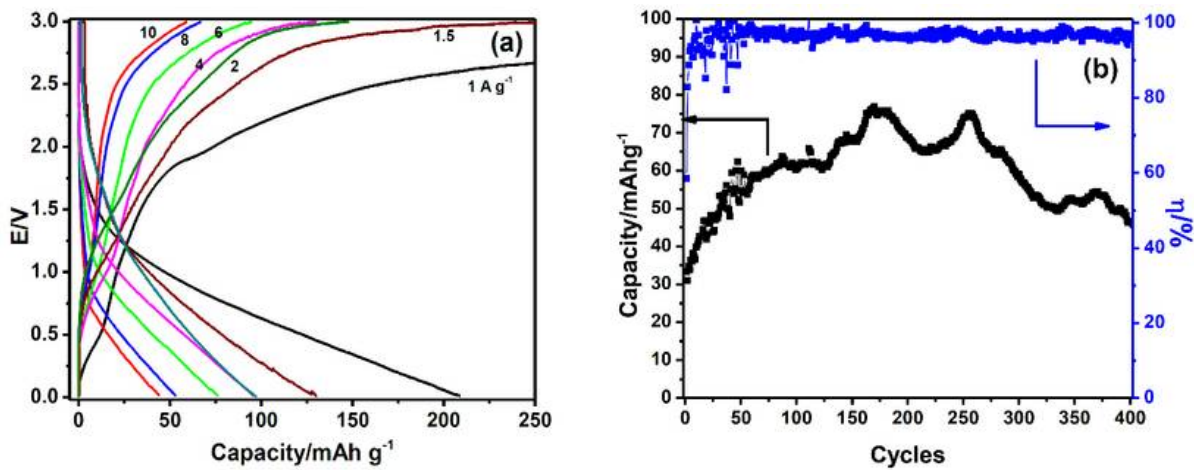

**Figure 4**. (a) Galvanostatic charge-discharge plots at different current densities (indicated) for sodium-ion capacitors prepared using Sn-gel polymer negative electrodes and biomass-based activated carbon positive electrodes with porous polymer film electrolyte, and (b) plots of capacity and Coulombic efficiency with respect to cycles for the same device.

To ascertain this claim, cycling of the NIC cells was carried out at constant current of 4 A g$^{-1}$ between 0 and 3V and the results are shown in Fig. 6.10 (b). The typical cell is able to maintain a large proportion (78.5 %) of its initial capacity even after 300 cycles. Not only this, another interesting phenomenon that is observed is that the capacity increases for the first few cycles before becoming stable. This is attributed to the restructuring of the electrode microstructure as seen in the FESEM images. Sn-particles break/split into smaller pieces upon repeated alloying/de-alloying.[23] These broken/split pieces are able to provide more surface area for access to the Na$^+$ ions; and the changes in the polymer structure that lead to the encasing of these broken particles leads to better transfer of charge. Both these effects ensure the improved performance of the NIC cells.

## 4 Conclusions

Flexible, solid-state sodium-ion capacitor (NIC) cells have been fabricated using a novel tin (Sn)-gel polymer negative electrode, activated carbon based positive electrode, and a porous polymer

(PVDFHFP) electrolyte film activated with liquid electrolyte $NaClO_4$-EC:PC. Sn-gel polymer electrode showed considerable changes in its microstructure after 30 charge-discharge cycles as evident from morphological and structural analyses. NIC made with Sn-gel polymer electrodes showed a high capacitance of ~210 mA g$^{-1}$ at current rate of 1 A g$^{-1}$. The specific energy and power were found to be 38.6 Wh kg$^{-1}$ and 666.8 W kg$^{-1}$, respectively. An important observation was the increase in specific capacity for the NIC for the first few cycles, attributed to the formation of the gel matrix which allows for amelioration of the effect of splitting of the Sn particles during alloying/de-alloying. The device showed stable cyclic performance for at least 175 cycles. The porous polymer film used in the NIC provided the required properties of safety, leakage-resistance, flexibility, and durability to the device.

## Acknowledgments

This work was supported by SERB (DST), New Delhi (Sanction No.: EMR/2016/002197). NY is grateful to University Grants Commission, New Delhi for providing Senior Research Fellowship.

## Conflicts of interest

The authors have no conflicts of interest to declare.